\documentstyle[sprocl]{article}

\def\be{\begin{equation}}
\def\ee{\end{equation}}
\def\bea{\begin{eqnarray}}
\def\eea{\end{eqnarray}}

\begin{document}

\title{\hfill OKHEP-98-10\\
JULIAN SCHWINGER AND THE CASIMIR EFFECT: THE REALITY OF ZERO-POINT
ENERGY}

\author{K. A. MILTON}

\address{Department of Physics and Astronomy, The University
of Oklahoma,\\Norman, OK 73019-0225 USA\\E-mail: milton@mail.nhn.ou.edu}


\maketitle\abstracts{ Julian Schwinger became interested in the
Casimir effect in 1975.  His original impetus was to understand
the quantum force between parallel plates without the concept
of zero point fluctuations of field quanta, in the language of
source theory.  He went on to consider applications to dielectrics
and to spherical geometries in 1977.  Although he published nothing
on the subject in the following decade, he did devote considerable
effort to understanding the connection between
acceleration and temperature in the mid
1980s.  During the last four years of his life, he became fascinated
with sonoluminescence, and proposed that the dynamical Casimir effect
could be responsible for the copious emission of photons by collapsing
air bubbles in water.}
\section{Introduction}
Julian Schwinger\footnotetext{Invited Talk at the Fourth Workshop
on Quantum Field Theory Under the Influence of External Conditions,
Leipzig, 14--18 September, 1998}
 was one of the pre-eminent theoretical physicists of
the 20th Century.  In addition to his famous solution of the problems of quantum
electrodynamics, he made many other contributions to science.  Nuclear
physics, the theory of angular momentum, a reformulation of quantum
kinematics, the quantum action principle, Euclidean field theory,
many-particle Green's functions, the electro-weak synthesis, non-Abelian
gauge theories, magnetic charge, source theory, and statistical models of atoms
are some of the major themes of his work.  (For a collection of some of
his papers on these and other topics, see Ref.~\cite{flato}.)
In this paper I document his contributions to the theory of the Casimir 
effect, an observable macroscopic consequence of quantum field theory, a
subject on which Schwinger worked to the day of his death.

\section{Casimir Effect}
In 1975 Schwinger became interested in the Casimir effect
through conversations with Seth Putterman.{\cite{seth}}
(Conversations with Walter Dittrich may have also played a 
role.{\cite{dittrich}})
The Casimir effect is a fundamental aspect of quantum field theory, indeed of
quantum mechanics, usually expressed as
an observable consequence of the zero-point
fluctuations of the normal modes of the electromagnetic field, or of
whatever quantum fields are relevant.
From another, complementary point of view, it is the macroscopic
manifestation of the van der Waals forces between the molecules
that make up material bodies.\footnote{Casimir and Polder in a heroic
calculation derived the retarded van der Waals force using nonrelativistic
quantum electrodynamics.{\cite{caspol}}  Shortly afterwards,
Niels Bohr asked Casimir what he was doing.  After hearing of this work, Bohr
`mumbled something about zero-point energy.  He gave me a simple 
approach.' {\cite{casimiratleipzig}}  That new approach was
reported in Paris, with a rederivation of the force between molecules,
and the force between a molecule and a conducting 
plane.{\cite{paris}}   We will discuss these results below.
The derivation of the force between conducting 
planes {\cite{casimir}} followed shortly.}
  Let us begin by reviewing the history
of this subtle yet fundamental phenomenon.

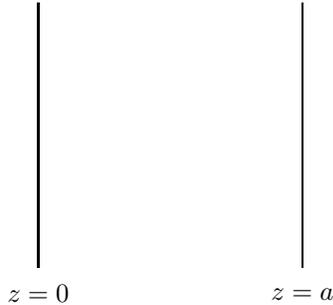
\begin{figure}
\centering
\begin{picture}(200,100)
\put(50,0){\line(0,1){100}}
\put(150,0){\line(0,1){100}}
\put(50,-10){\makebox(0,0){$z=0$}}
\put(150,-10){\makebox(0,0){$z=a$}}
\end{picture}
\caption{Parallel conducting plates separated by a distance $a$.}
\label{fig16.1}
\end{figure}

In 1948 H. B. G. Casimir {\cite{casimir}} considered
two perfectly conducting parallel plates in vacuum separated by a distance
$a$.  See Figure \ref{fig16.1}.
Although it is usually, and correctly, asserted that the zero-point
oscillations of the fields in the vacuum are unobservable, the presence
of the boundaries changes that situation.  On a perfect conductor,
the tangential component of the electric field must be zero.  Casimir
observed that one could measure the {\em difference\/} between the
zero-point energy in vacuum and the zero-point energy in the presence
of the boundaries,
\begin{equation}
E_c=\sum\textstyle{1\over2}\hbar\omega_{\rm cond}
-\sum\textstyle{1\over2}\hbar\omega_{\rm vac},
\end{equation}
where the subscripts refer to the normal modes, of frequency $\omega$,
 in the presence of the
conducting boundaries, and in the vacuum of unbounded space, respectively,
and the sums are over all possible normal modes in the two situations.
Of course, this formula is purely symbolic, since both sums are horribly
divergent.  In effect, Casimir gave a proper definition to this sum,
and was able to extract a finite result, the Casimir energy for
this geometry.  Because the plates are assumed to extend indefinitely
in the transverse directions (the $x$ and $y$ directions in the coordinate
system shown in Fig.\ \ref{fig16.1}), it is the energy $E_c$ per unit area of
the plates which Casimir calculated,
\begin{equation}
E_c=-{\pi^2\over720}{\hbar c\over a^3},
\label{casen}
\end{equation}
or, by differentiating with respect to the plate separation $a$, the force
per unit area $f$,
\begin{equation}
f=-{\pi^2\over240}{\hbar c\over a^4} = -0.013{1\over a^4} \mbox{dyn
($\mu$m)$^4$/cm$^2$}.
\label{casf}
\end{equation}
Note the minus sign in these expressions.  That means that the Casimir
force is attractive: The two plates are pulled toward each other.  To
this day, the sign of this effect defies intuitive understanding.

As early as 1956 experimental results appeared more or less in agreement
with the Casimir theory.{\cite{casexp}}  It is very difficult
to measure the Casimir force between conductors, because any small
stray electrical charge on the conductors will give rise to a much
larger electrical attraction or repulsion.  
(Very recently, however, definitive experimental results, completely
consistent with Casimir's prediction, have been 
published.{\cite{prl}})  On the other hand, it was clear from
the outset that there was nothing particularly special about having
conducting boundaries.  Starting in 1955, E. M. Lifshitz and collaborators
in Moscow developed the theory of the Casimir effect for parallel
dielectrics, that is for boundaries as shown in Fig.\ \ref{fig16.1},
but with different dielectric constants $\epsilon(z)$ in the three regions,
\begin{equation}
\epsilon(z)=\left\{\begin{array}{cc}
z<0:\quad&\epsilon_1\\
0<z<a:\quad&\epsilon_3\\
a<z:\quad&\epsilon_2\end{array}\right..
\end{equation}
For this geometry Lifshitz et al.\ obtained a somewhat complicated
formula {\cite{lifshitz}} that expressed the force per unit
area between the dielectrics in terms of the dielectric constants,
which were assumed to be functions of the frequency:
\begin{eqnarray}
f&=&-{1\over8\pi^2}\int_0^\infty d\zeta\int_0^\infty dk^2 2\kappa_3
\Bigg\{{1\over{\kappa_3+\kappa_1\over\kappa_3-\kappa_1}
{\kappa_3+\kappa_2\over\kappa_3-\kappa_2}e^{2\kappa_3a}-1}
\nonumber\\
&&\qquad\mbox{}+{1\over{\kappa'_3+\kappa'_1\over\kappa'_3-\kappa'_1}
{\kappa'_3+\kappa'_2\over\kappa'_3-\kappa'_2}e^{2\kappa_3a}-1}\Bigg\},
\end{eqnarray}
where $\zeta={1\over i}\omega$ is the imaginary frequency, 
$k^2={\bf k}_\perp^2$ is the square of the transverse momentum,
$\kappa^2=k^2+\zeta^2\epsilon$, and $\kappa'=\kappa/\epsilon$.
 In fact, this
Lifshitz force was confirmed, based on knowledge of these dielectric
constants, in a beautiful experiment of Sabisky and 
Anderson,{\cite{sabisky}} who measured the force holding
a film of superfluid liquid helium to a SrF$_2$ substrate, to high
precision, over distance scales differing by a factor of 1000.  So by 1973 there
was no doubt of the theoretical or experimental reality of the Casimir
effect.\footnote{Schwinger noted in a talk in 1988 in honor of
Herman Feshbach that Leonard Schiff had proposed that van der Waals forces
were responsible for holding such a helium film to a surface, thus
anticipating the Lifshitz theory.{\cite{archive}}} 

As noted, in 1975, Schwinger became interested in explaining the
Casimir effect in the source theory language, which `makes no reference
to quantum oscillators and their associated zero point energy.' \cite{174}
As usual, his presentation was first to his  field theory class,
and only then did he write a short publication.\cite{174}  
Anticipating that the effect of
the two polarizations of electromagnetism was merely a doubling
of that for a single, massless, scalar mode, his derivation consisted, first,
in obtaining the general expression for the infinitesimal change in
the action under an infinitesimal change in the physical parameters,
\begin{equation}
\delta W={i\over2}\int (dx)(dx')D(x,x')\delta D^{-1}(x',x),
\label{delW}
\end{equation}
where $D$ is the massless propagation function or Green's function,
or the equivalent change in the energy
\begin{equation}
\delta{\cal E}=-{i\over2}\int (d{\bf r})(d{\bf r}')d\tau
D({\bf r,r'},\tau)\delta D^{-1}({\bf r',r},-\tau),
\end{equation}
which ignores transient effects. 
 Then, by inserting an appropriate 
Green's function that satisfies the Dirichlet boundary conditions at
$z=0,a$, written in terms of the longitudinal eigenfunctions,
$\sqrt{2/a}\sin(n\pi z/a)$, he obtained the following
formula for the change in the energy per unit area if the separation
is changed by an amount $\delta a$, due to the Green's functions
in the region $0<z<a$:
\begin{equation}
{\delta{\cal E}_a\over A}={1\over4\pi}{\delta a\over a}{1\over i\tau}
{d^2\over d\tau^2}{1\over1-e^{-i(\pi/a)\tau}},
\label{in}
\end{equation}
where the limit $\tau\to0$ is understood.  This result is divergent
in that limit.  But Schwinger then subtracted off the contribution
from the region on the other side of the plate,
$a<z<L$ (an additional conducting plate is placed
at $z=L\gg a$),
 which may be immediately inferred from Eq.~(\ref{in}) to be
\begin{equation}
{\delta{\cal E}_{L-a}\over A}=-{1\over 4\pi}\delta a{1\over i\tau}
{d^2\over d\tau^2}{1\over\pi i\tau}.
\label{out}
\end{equation}
The force per unit area is then immediately found 
from the sum of Eq.~(\ref{in}) and Eq.~(\ref{out}) to be
\begin{equation}
f=-{1\over A}{\partial{\cal E}\over\partial a}=-{\pi^2\over480}{1\over a^4},
\end{equation}
indeed, exactly one-half Casimir's result (\ref{casf}).

Schwinger concluded this note by rederiving the effect of finite
temperature, in particular, the high-temperature limit,
\begin{equation}
kT\gg{\pi\over a}:\quad f_T=-{\zeta(3)\over8\pi}{kT\over a^3},
\end{equation}
which had first been obtained by F. Sauer and J. 
Mehra.{\cite{sm}}  Schwinger justified this publication,
apart from it giving the Casimir effect
 a source theory context free from an operator
substructure, by quoting from C. R. Hargreaves,{\cite{har}}
who stated that `it may yet be desirable that the whole general theory
be reexamined and perhaps set up anew.'  The context of the latter remark
was a discrepancy between the temperature dependence
found between conducting plates, and that found from
the temperature-dependent Lifshitz formula {\cite{lifshitz}}
when the dielectric constant in the region outside $0<z<a$ is set
equal to infinity, a process which should correspond to a perfect
conductor.  Unbeknownst to Schwinger, this error had been corrected
subsequently.{\cite{lifcorr}} (Hargreaves had corrected
another error in Lifshitz' paper having to do with the effect of
imperfect conductors.)

It was partly this (nonexistent) discrepancy, but primarily the challenge
to understanding the phenomenon in his own language, that led Schwinger,
and his 
postdocs Milton and DeRaad, to write `Casimir Effect in Dielectrics,' \cite{187}
in which the Lifshitz formula for the Casimir force between parallel
dielectrics was rederived in an elegant, action-principle based, Green's
function technique.  The key point here was that the effective product
of electric fields could be represented in terms of the classical
electromagnetic Green's dyadic,
\begin{equation}
{\bf E(r)E(r')}\bigg|_{\rm eff}
={\hbar\over i}\mbox{\boldmath{$\Gamma$}}({\bf r,r'};\omega),
\end{equation}
where the Green's dyadic satisfies
\begin{equation}
-\mbox{\boldmath{$\nabla\times(\nabla\times\Gamma$}})+\omega^2\epsilon
\mbox{\boldmath{$\Gamma$}}=-\omega^2\mbox{\boldmath{$1$}}\delta({\bf r-r'}).
\end{equation}
From $\mbox{\boldmath{$\Gamma$}}$ Schwinger, Milton, and DeRaad calculated
the change in the energy, using a method similar to that
sketched above, or equivalently, the force directly from the electromagnetic
stress tensor,
\begin{equation}
T_{zz}={1\over2}[H_\perp^2-H_z^2+\epsilon(E_\perp^2-E_z^2)],
\end{equation}
where $\bf H$ is calculated from $\bf E$
(and hence $\mbox{\boldmath{$\Gamma$}}$) using Maxwell's equations.
Removing constant divergent terms from the result,
 the so-called volume stress,
which would be present if a given dielectric extended over all space,
they succeeded in rederiving the Lifshitz formula.  As a special case,
they took the perfect conductor limit noted above ($\epsilon\to\infty$ in
the external region) and obtained the Casimir result, as well as the
appropriate high and low temperature limits found by Sauer and
Mehra.{\cite{sm}}  They also showed how, in the case
of tenuous dielectrics, i.e., in the case when $\epsilon-1\ll1$,
the Casimir force could be thought of as the superposition of the
van der Waals attractions between the individual molecules 
(separated by a distance $r$) that made up the media,
\begin{eqnarray}
\mbox{large separations:}\quad V&=&-{23\over4\pi}{\alpha_1\alpha_2\over r^7},
\\
\mbox{small separations:}\quad V&=&-{3\over\pi r^6}\int_0^\infty
d\zeta\alpha_1(\zeta)\alpha_2(\zeta),
\end{eqnarray}
where $\alpha=(\epsilon-1)/4\pi N$ is the electric polarizability
of the molecules, with number density $N$.
These are the van der Waals potentials originally derived by Casimir
and Polder{\cite{caspol}} and by Fritz 
London,{\cite{london}} respectively.

These results were all explicitly contained in the much earlier papers
by Lifshitz and collaborators,{\cite{lifshitz}} to
whom due acknowledgement was made.  Nevertheless, Lifshitz was somewhat
offended by this paper, and he wrote Schwinger a letter: 
`Thank you for the preprint of your \dots paper \dots.  It was gratifying
to know of your interest in my earlier work.'

`Of course, the method
adopted in this paper is far superior than [sic] the method which was
used in my first paper of 1954.  But it seems to me that it is almost
identical with the method developed later by I. Dzyaloshinskii, L. P.
Pitajevskii, and myself.  The derivation of my results by this method
was published in our joint paper in {\it Advances of Physics}, 1961 (identical
with the paper in {\it Soviet Physics, Uspechi}, referred to in your preprint);
it was also reproduced in the book by Abrikosov, Gorkov, Dzyaloshinskii
on the {\it Field Theoretical Methods in Statistical Physics\/} (English
translation, Prentice Hall, 1963).'

`As to the formula for the low temperature limit of the force between
the two perfect metallic surfaces (formula 3.17 of your paper), the error
in sign in my paper was the result merely of an unfortunate slip in
rewriting the Euler summation formula, and not of a deeper origin.  This error
has since [been] noticed by different authors both in our country and 
elsewhere.' {\cite{lifcorr}}

The only really new result in this paper was an attempt to derive the
surface tension for an ideal liquid (liquid helium) from such considerations,
by examining the effect of a change of shape of the surface on 
the energy.  `The second-order change in the energy \dots is directly
related to the surface tension.'~\cite{187} Unfortunately, a quadratically
divergent result was obtained.  However, with reasonable numbers inserted
to provide a physical cutoff to the divergence,
a value for the surface tension, and for the latent heat, could be obtained
crudely in agreement with the observed values
to within a factor of two or three.  This idea remains
provocative yet unresolved.

A few months later the same three authors wrote a second paper on
Casimir phenomena, entitled `Casimir Self-Stress on a
Perfectly Conducting Spherical Shell.' \cite{188}  The impetus for this
work went back to another paper of Casimir, this one in 1956, in which
he suggested that the attractive Casimir force could balance the Coulomb
repulsion of a semiclassical model of an 
electron.{\cite{casimir2}}  More precisely, it
had long been known that a purely electromagnetic classical model of
an electron was impossible, that one had to add the so-called Poincar\'e
stresses to stabilize the particle.  Casimir now suggested that those
stresses could arise from quantum mechanics.  Indeed, if a reasonable
guess extrapolated from the parallel plate calculation was used, one
could calculate a value for the charge on the electron, or better, the
fine structure constant, $\alpha=e^2/\hbar c$, consistent, perhaps, with
the experimental value, $\alpha=(137.036\dots)^{-1}$.

It remained for Timothy Boyer, a student of Sheldon Glashow at Harvard, to
take up the challenge of a real calculation for the spherical geometry
in 1965.  He calculated the change in the zero-point energy due to the
presence of a perfectly conducting spherical shell of radius $a$.
Both modes interior to and exterior to the shell had to be included
in order to get a finite result.  This impressive calculation was difficult and
subtle, and involved extensive numerical calculation.  His result,
obtained after three years of work, was accurate to only one significant
figure, but it was of the {\em opposite\/} sign compared to the one found
by Casimir in the parallel geometry: {\cite{boyer}}
\begin{equation}
E_B=+{0.9\over 2a}.
\end{equation}
His expression was subsequently evaluated more accurately, to three significant
figures, by Davies.{\cite{davies}}

Because this result was so surprising, and devastating to Casimir's
electron model, it was an obvious target for a recalculation by
Schwinger and his postdocs, now that their improved Green's function
machinery had been honed.  By the end of 1977 they had derived a
compact formula for the Casimir energy of a conducting shell, much
simpler than that of Boyer,
\begin{equation}
E=-{1\over2\pi a}\sum_{l=1}^\infty(2l+1){1\over2}\int_{-\infty}^\infty
dy\,e^{i\epsilon y}x{d\over dx}\log(1-\lambda_l^2),
\label{mds}
\end{equation}
where the sum is taken over the different angular momentum modes, the
integral is over (imaginary) frequencies, $y={1\over i}\omega a$, the quantity
$x=|y|$, and the logarithm depends on
\begin{equation}
\lambda_l(x)=(s_le_l)'(x)
\end{equation}
(where the prime denotes differentiation). The functions $e_l$ and
$s_l$ are given in terms of modified Bessel functions,
\begin{eqnarray}
s_l(x)&=&\sqrt{\pi x\over2}I_{l+1/2}(x),\\
e_l(x)&=&\sqrt{2 x\over\pi}K_{l+1/2}(x).
\end{eqnarray}
The expression (\ref{mds}), which is formally divergent, has been regulated
by evaluating the underlying Green's function at unequal times, $t=t'+\tau$,
i.e., by `time-splitting.'  At the end of the calculation one is to 
take the limit $\epsilon=\tau/a\to0$.
Unfortunately, at this point Milton and DeRaad had a bit of difficulty
in seeing how to extract a number from this formula, so a few months
passed.  (Schwinger had contented himself with deriving the formula.)
Unfortunate, because just at that point a paper by Balian and Duplantier
appeared,{\cite{baldup}} who obtained a different formula,
based on a multiple scattering formalism, and obtained a result, consistent
with Boyer's number,  but now accurate to three significant figures.
So the postdocs worked hard, discovered how to extract a reliable
answer based on the use of uniform asymptotic approximations
(the first term of which was accurate to 2\%, while Balian and Duplantier's
first approximation was only accurate to 8\%), and obtained the result
accurate to {\it five\/} significant figures,
\begin{equation}
E={0.92353\underline{1}\over2a}.
\end{equation}

The reaction from Boyer and Balian was rather unexpected.  In a letter
to Lester DeRaad (DeRaad and Boyer, of course, had been fellow graduate students
at Harvard) Tim Boyer wrote,
`The calculations presented seem sophisticated, and presumably are
carefully done.  However, the comments on my work in the text of the
Casimir sphere paper are hardly generous; my colleagues would characterize
them differently.' \cite{boyerlet}

He went on to apprise DeRaad of the Davies calculation, and to give further
experimental references, which were incorporated into the published
papers. In addition, an appreciative comment about Boyer's work was
inserted into Ref.~\cite{188}.

Roger Balian wrote Schwinger to say
`I guess it would be interesting to compare our respective approaches, 
which have the common feature of being based on the elimination of fields
and consideration of sources.  Our formalism was mainly intended to deal
with arbitrary geometries;  it is based on an expansion which converges
rapidly in cases of interest (slightly deformed conducting sheet, spherical
shell, etc.\dots).  However, we construct the electric Green's function in
terms of fictitious monopole currents, and restrict to conductors.  Your
approach has the advantage of allowing the treatment of dielectrics; I do
not see, however, how to use it for arbitrary geometries; on the other 
hand, would you obtain instabilities of the surface of a dielectric at
$T\ne0$, thus generalizing the effect which we pointed out for a conducting
foil?' {\cite{balianlet}}
Since this letter was dated December 28, 1977, more than five months before
Schwinger's paper on the Casimir effect for a sphere was submitted,\cite{188}
it seems likely that at that point Balian had only seen the dielectric 
paper; \cite{187}
hence the remark about geometries.

Milton responded to both of these letters graciously, and promised to look
at Balian's technical points in the future, but that never occurred.

Schwinger's papers on the Casimir effect were influential, not for
their explicit results, which, as we have seen, were mostly well-known,
but for the development of powerful techniques of attacking such
problems, which continue to be exploited.  A recent example is the
study of the dimensional dependence of the Casimir effect in hyperspheres
by Bender and Milton.{\cite{dimcas}}

\section{Acceleration and Temperature}

In the mid 1980s Schwinger received an honor,
the Monie Ferst Medal given by the Georgia Institute
of Technology chapter of Sigma Xi.  The associated Symposium, held
on May 20, 1986, consisted of technical talks by three of his former
students, Milton, Ken Johnson, and Margaret Kivelson, and a provocative
talk by Schwinger on `Accelerated Observers and the Thermal Power
Spectrum of the Vacuum.'  This talk reported his work on the `Unruh'
effect.\cite{unruh}\footnote{This
interest may have been sparked by a 1983 letter from Kirk McDonald of
Princeton, on the Unruh effect.{\cite{archive}}}
  Although
he had spent considerable time working on this project, and presented
a most interesting presentation on the subject, he did not then, or
later, ever write up this work.  All that was printed is his abstract for the
symposium:
`Source theory, with its foundation in idealizations
of particle emitters and absorbers (detectors), provides a natural,
self-con\-tained approach that is intermediate between the mathematical
attitudes of quantum field theorists and the physical consideration of
specific detection mechanisms.  The periodicity inherent in the circular
coordinate form of the Euclidean Green's function, as transformed into
hyperbolic (Rindler) coordinates, immediately yields the characteristic 
property of a thermal Green's function.  The explicitness with which this
can be done assists in recognizing that, despite the thermal nature of
the spectrum, there are definite phase relations that would show up in
other experiments.' \cite{sigxi}
 
The Schwinger Collection at UCLA does possess a few manuscripts on the
subject of acceleration and radiation
which Schwinger started but did not complete.{\cite{archive}} 
 There is also a draft of a paper with Manuel Villasante,
dated 1994, entitled `Acceleration, Black Holes, and Temperature;'
this paper was never submitted, presumably because of Schwinger's 
illness.\footnote{Schwinger died of pancreatic cancer in July 1994, having
been diagnosed in February of that year.}
Villasante received his Ph.D. under Robert Finkelstein's direction,
and became Schwinger's final postdoc.  However, they never
completed a paper together. The first part of the extant 56 page manuscript
is essentially equivalent to the earlier solo effort of Schwinger
on accelerated detectors; Villasante's contribution largely consisted
of extending the ideas to a Schwarzchild space, hoping thereby to make the
connection with Hawking radiation.\cite{hawking}  
In fact, Villasante recalls they only had
one brief conversation about the work, late at night, and Schwinger
never responded to his messages.{\cite{villa}}\footnote{Schwinger also had a
Greek student in his last years, Evangelos Karagiannis, who did his
Ph.D on a related topic,\cite{kara}
but `he also found it impossible to get hold of Schwinger for 
anything.'~\cite{villa}
What these late collaborators failed to appreciate was that `Schwinger was
hard to work with if you wanted guidance, but easy to work with if you
wanted inspiration.'~\cite{arnowitt}}
Schwinger apparently never felt this work was complete 
enough.{\cite{fink}}

\section{Casimir effect and sonoluminescence}

Schwinger's last physics endeavor marked a return to the Casimir effect,
of which he had been enamored nearly two decades earlier.  It was sparked by
the remarkable discovery of single-bubble sonoluminescence.  It was
not coincidental that the leading laboratory investigating this phenomenon
was, and is, at UCLA, led by erstwhile theorist Seth Putterman, long
a friend and confidant.  Putterman and Schwinger shared many interests
in common, including appreciation of fine wines, and they shared a similar
iconoclastic view of the decline of physics.  So, of course, Schwinger
heard about this remarkable phenomenon from the horse's mouth, and was
greatly intrigued.

What is sonoluminescence?  The word means the conversion of sound into
light.  As such, it had been observed since the thirties,{
\cite{sono}} but this so-called multiple bubble sonoluminescence was
hardly investigated, and was nearly completely forgotten by the last decade
of this century.  Not completely, because Tom Erber, on one of his many
visits to UCLA, told Putterman about this old 
effect,{\cite{seth}} and, in short order,
a much more remarkable version of the effect was discovered.
If a single bubble of air is injected into a beaker of water, and
held in a node of a standing acoustic wave set up in the water, 
the bubble will begin to expand and contract in concert with the frequency
of the standing wave.  If the ultrasonic wave has a frequency of about
20,000 Hertz, and a pressure amplitude of about one atmosphere, a small
suspended bubble of air will expand and collapse 20,000 times a second,
undergoing a change in radius of a factor of 10 or more
 (and hence, in volume, of at least 
a factor of 1,000), from, say $4\times10^{-3}$ cm to $4\times10^{-4}$ cm.
If the parameters are chosen just right (including a small percentage
of noble gas, for example, the amount of argon in our atmosphere, seems
essential), exactly at minimum radius a
bright flash of light is released from the bubble.  This flash of light
consists of approximately one million optical photons, so that about 10 MeV
of energy is converted into light on each collapse.  This flash of light,
integrated over many cycles, is bright enough to be visible to the  naked
eye if the water is observed in a darkened room.  (The author has seen
this effect for himself.) Whatever produces the flash of light is
sufficiently non-catastrophic that it does not in any way disrupt the
bubble, and the periodic collapse and re-expansion continues for many
minutes, perhaps months. For a review of the experimental situation,
see Ref.~\cite{putt}.

The hydrodynamics of the bubble collapse and re-expansion appears to be quite
well understood.  What is not understood at all is how some of the
energy in the bubble, extracted from the sound field, is converted
into the intense flash of light.  The duration of the flash has not
been determined, but it is less that $10^{-11}$ seconds, much smaller
than the period of the bubble collapse, but apparently long compared
to the period of optical photons (about $10^{-15}$ seconds).
Although there have been various classical and quantum hypotheses put
forward, they tend not to be, in the words of Putterman, 
`falsifiable.' {\cite{putt}}

Of course, Putterman told Schwinger about the phenomenon right away.
He called Schwinger at home, and immediately Schwinger drove down to
see it.  At first Schwinger had difficulty in seeing the faintly glowing
bubble. Putterman told him to `look at $r=0$,' and soon he saw the
bubble at the center of the spherical vessel.
Schwinger's reaction was `I'm shaken.'  He at once started work
on the problem of understanding what was happening.\cite{seth}

Schwinger immediately had the idea that
a dynamical version of the Casimir effect might play a key role.
In a letter to Putterman `Re: nanosecond sonoluminescence' wherein
he proposes the Casimir effect mechanism, Schwinger opens with a
quotation, presumably written on Martin Luther King Day:
`MLK: ``I have a dream.''  JS: ``I have a 
feeling.'''~{\cite{archive}}
The idea was that the virtual photons present, due to the Casimir
effect, or in conventional language, vacuum fluctuations, in a bubble
in a dielectric medium could be converted into real photons because
the radius of the bubble is rapidly changing.  This is, in fact, presumably
closely related to the Unruh effect~{\cite{unruh}}
in which a moving mirror radiates a black body spectrum of photons---in
turn closely allied with Hawking radiation from a black hole.\cite{hawking}
(Recall that Schwinger had worked for a while on the Unruh effect
in the mid 1980s into the 1990s, although he never completed a
paper on the subject.\footnote{In 1990, just before single bubble 
sonoluminescence was discovered, Schwinger wrote a manuscript entitled 
`Superluminal Light' (a later version was called `Tachyonic Light')
which was a reaction to the claim by K. Scharnhorst and
G. Barton that light speeds greater than the speed of light in vacuum
are possible in a parallel plate capacitor, the original Casimir effect
geometry, indeed as an induced consequence of the Casimir 
effect.{\cite{barton}}
  Unlike them, Schwinger found the effect was nonuniform,
dispersive (that is, frequency dependent), and that the effect persisted
if only a single plate was present.{\cite{archive}}}) So there
were two challenges for Schwinger.  One was to develop the `dynamical
Casimir effect' for the spherical geometry of a bubble, and the second
was to apply that effect to the hydrodynamic situation of a collapsing
bubble in sonoluminescence.  

The first step was initiated\footnote{Early in the process he gave the talk
`A Progress Report: Energy Transfer in Cold Fusion and Sonoluminescence.'
The title here hearkens back to Oppenheimer's seminar given when Schwinger
first came to Berkeley in 1939.{\cite{archive}}}
 through two papers Schwinger published
in {\it Letters in Mathematical Physics}, edited by the frequent UCLA visitor
Mosh\'e Flato, as sequels to his first, 1975, Casimir
paper,\cite{174} also published in the same journal.  
In the first,\cite{221} he derived the original Casimir effect for
parallel conducting plates by an elegant proper-time approach,
while in the second,\cite{222} he reconsidered dielectric slabs. In both cases,
the emphasis was on energy rather than force.
He followed\footnote{Actually Ref.~\cite{223} 
was submitted essentially simultaneously
with Ref.~\cite{222}.}
 this by two somewhat longer articles in the {\it Proceedings of
the U. S. National 
Academy of Sciences}.  (Unlike Feynman,{\cite{feynbio}}
Schwinger continued throughout his career
 to find the Academy a useful scientific venue.)
 In the first, he rederived, for the third time,
  the Lifshitz theory for the Casimir
effect between parallel dielectric slabs,\cite{223} in an efficient way
making use of an explicit break-up into Transverse Electric (TE)
and Transverse Magnetic (TM)
 modes.  As had been done in his earlier collaborative work,\cite{187}
he explicitly removed volume and surface energies:
`One finds contributions to $E$ [the energy] that, for example, are
proportional \dots to the volume enclosed between the slabs.  The
implied constant energy density---independent of the separation of the 
slabs---violates the normalization of the vacuum energy density to zero.
Accordingly, the additive constant has a piece that maintains the vacuum
energy normalization.  There is also a contribution to $E$ that is
proportional to [the area], energy associated with individual slabs.
The normalization to zero of the energy of an isolated slab is maintained
by another part of the additive constant.' \cite{223}
Then he turned to the case of interest for sonoluminescence, spherical
dielectrics.  In `Casimir Energy for Dielectrics: Spherical Geometry'~\cite{224}
he began an elegant treatment of the Casimir effect in that situation.
Unfortunately, he only treated the TE modes, and went only far enough
to see that the parallel geometry result is recovered if a careful 
limit of the radius of the sphere going to infinity
is taken.  Explicitly, he left the details to Harold!\footnote{Harold,
introduced as an acronym  for the `hypothetical alert 
reader of limitless dedication'
in Ref.~\cite{sagredo}, was the name of Schwinger's older brother.}

But Harold, or Sagredo,{\cite{sagredo}}
 had been over this ground already.  Thirteen
years earlier, while still at UCLA, Milton had computed the Casimir
effect for a dielectric ball.{\cite{kim}}  Perhaps
Schwinger can be forgiven his ignorance of his former student and
postdoc's work by the fact that this paper was completed and published
after Milton had gone to Ohio State.\footnote{However, in Milton's
last meeting with Schwinger, in December 1993, Schwinger did not
wish to be reminded of this earlier work.}  In any case, Schwinger did
not get far enough with this calculation to apply it to sonoluminescence.
Instead, when he started to develop his theory of sonoluminescence
in a series of five papers\footnote{There are notes for at least three further
papers in Schwinger's files on `Casimir Light,' the last being
subtitled `A Study in Green.'  These must represent his last scientific 
work.{\cite{archive}}}
 in the {\it Proceedings of the U. S. National 
Academy of Sciences\/}\cite{225,230} he simply wrote down a naive approximation
for the Casimir energy obtained, in effect, by subtracting the zero-point
energy of the vacuum from that for the medium, giving the quartically
divergent formula,
\begin{equation}
E_{\rm bulk}={4\pi a^3\over3}\int{(d{\bf k})\over(2\pi)^3}{1\over2}k
\left(1-{1\over n}\right),
\end{equation}
where $n$ is the index of refraction of the medium.  Schwinger had
forgotten his own injunction of subtracting off the volume energy,
that term which `would be present if either medium filled all space.'
Since this expression is very divergent, it is extraordinarily sensitive
to the cutoff which must be used on physical grounds to give a finite
result.  However, if a plausible ultraviolet cutoff is used, 
Schwinger obtained a sufficiently large Casimir energy, $E_{\rm bulk}\sim
10$ MeV.

The problem is that the bulk energy Schwinger considered is not
relevant to sonoluminescence.{\cite{kim}}  
It is, in fact, a kind of self energy, one that contributes to the density
of the water, and of the gas, that is already phenomenologically
described.  As further noted above, the same is true of the surface
energy, it being subsumed into the definition of the surface tension.
The correct conclusion from the calculation of Refs.\ \cite{kim} and
\cite{brevikandbarton} is that the Casimir energy is very small,
\begin{equation}
E_c={23(n-1)^2\over384 a},
\end{equation}
which only amounts to about $10^{-3}$ eV in the case of sonoluminescing
bubbles (as well as having the wrong sign), 
and therefore is completely irrelevant to sonoluminescence.\footnote{Schwinger
did treat briefly the production of photons through the instantaneous
collapse of the bubble.  This will be discussed in another contribution
to this Proceedings.\cite{Proc2}}

Schwinger's final paper,
 on sonoluminescence,\cite{230} was published in the month of his death.
As we noted he was typically unaware of some
of his colleagues' own papers relevant to the subject, 
but, atypically, he  was
very explicitly seeking Milton's collaboration in the last year of his life
(Milton talked to him at some length in December 1993, at the annual
Christmas party given by the Alfredo Ba\~nos', which he and Clarice often
attended,\footnote{His role was to hide the three kings in the Christmas
tree.}
 and at a subsequent lunch). He also arranged to have the earlier papers,
 on Casimir energy, but not the later ones, on sonoluminescence, sent to
 Milton.
 Schwinger felt that `carrying out
that program is---as one television advertiser puts it---job one.'~\cite{229}  
It seems apparent that he was aware of the inadequacies of
his treatment of the Casimir effect, and was looking for additional
expertise and strength.  The subject is not completely closed, 
because there are serious subtleties in these Casimir calculations,
the adiabatic approximation (that is, treating the bubble radius as
slowly varying on an electromagnetic time scale)
may be invalid, and most likely a
shock forms, which allows for discontinuities on very short time
scales.  So Schwinger's ideas here are still being explored.\cite{eberviss}
Perhaps Julian Schwinger will ultimately have the last laugh!

\section{Conclusions}

As we have noted, Schwinger explicitly and implicitly drew parallels
between cold fusion and sonoluminescence.  At first blush this seems
implausible.  After all, sonoluminescence without doubt exists, while
cold fusion does not.  But Schwinger's point was one of overcoming
seemingly impossibly different scales.  In the case of cold fusion,
how can the Coulomb barrier be overcome at very low energies; in the
case of sonoluminescence, how could hydrodynamics, characterized by
acoustic phonons, couple to quantum electrodynamics, characterized
by much higher energy photons?  It is natural that he would find
the attempt to solve these conundrums challenging.  And, as it became
increasingly untenable to pursue cold fusion, he shifted his efforts
toward the experimentally confirmed sonoluminescence.

Seth Putterman recounts his final meeting with Schwinger two days
before his death.  Schwinger did not want to talk about history, but
about physics, and wanted to know what was new in sonoluminescence.
Putterman told him of the puzzling fact that water is the `friendliest'
liquid for the phenomenon, and that the effect only appears if about
1\% noble gas is present.  Schwinger thought for a bit, and said,
`It probably has something to do with 
evolution.'~{\cite{seth}}  Heady stuff indeed!

\section*{Acknowledgments}
I thank Michael Bordag for inviting me to present this material at
the Fourth Workshop on Quantum Field Theory Under the Influence of
External Conditions.  I thank him, H.B.G. Casimir, Iver Brevik, Gabriel
Barton, and Walter Dittrich for useful conversations there.  This work
was supported in part by a grant from the U.S. Department of Energy.


\section*{References}
\begin{thebibliography}{99}

\bibitem{flato} {\it Selected Papers (1937--1976) of Julian Schwinger},
ed.~M. Flato, C. Fronsdal, and K. A. Milton (Reidel, Dordrecht, 1979).

\bibitem{seth} Interview with Seth Putterman by K. Milton, July 1997.

\bibitem{dittrich} Letter from Walter Dittrich to K. Milton, September 1998.

\bibitem{caspol} H. B. G. Casimir and D. Polder, {\it Phys.\ Rev.} {\bf 73}, 
360 (1948).

\bibitem{casimiratleipzig} H. B. G. Casimir, talk at the Fourth Workshop
on Quantum Field Theory Under the Influence of
 External Conditions, Leipzig, September 1998.

\bibitem{paris} H. B. G. Casimir, {\it Colloq. sur la Theorie de la Liaison
Chemique}, Paris, 12--14 April, 1948.

\bibitem{casimir} H. B. G. Casimir, {\it Proc.\ Kon.\ Ned.\ Akad.\ Wetensch.}
{\bf 51}, 793 (1948).

\bibitem{casexp} B. V. Deriagin and I. I. Abrikosova, {\it 
Zh.\ Eksp.\ Theor.\ Fiz.}
{\bf 30}, 993 (1956); {\bf 31}, 3 (1956) [Engl. transl. {\it Soviet Phys. 
JETP\/} {\bf 3}, 819 (1956); {\bf 4}, 2 (1957)]; A. Kitchener and A. P. Prosser,
{\it Proc.\ Roy.\ Soc.\ (London)} A {\bf 242}, 403 (1957); W. Black, J. G. V.
de Jongh, J. Th. G. Overbeck, and M. J. Sparnaay, {\it Trans.\ Faraday Soc.}
{\bf 56}, 1597 (1960); A. van Silfhout, {\it Proc.\ Kon.\ Ned.\ Acad.\ Wetensch.}
B {\bf 69}, 501 (1966); R. H. S. Winterton, {\it Contemp.\ Phys.} {\bf 11}, 559
(1970); J. N. Israelachivili and D. Tabor, {\it Proc.\ Roy.\ Soc.\ (London)} 
A {\bf 331}, 19 (1972).

\bibitem{prl} S. K. Lamoreaux, {\it Phys.\ Rev.\ Lett.} {\bf 78}, 6 (1997);
U. Mohideen and A. Roy, physics/9805038, to be published in {\it Phys.\
Rev.\ Lett.}

\bibitem{lifshitz} E. M. Lifshitz, {\it Zh.\ Eksp.\ Teor.\ Fiz.} {\bf 29}, 94 
(1955) [Engl.\ transl.: {\it Soviet Phys.\ JETP\/} {\bf 2}, 73 (1956)];
I. D. Dzyaloshinskii, E. M. Lifshitz, and L. P. Pitaevskii, 
{\it Usp.\ Fiz.\ Nauk\/}
{\bf 73}, 381 (1961) [Engl.\ transl.: {\it Soviet Phys.\ Usp.} 
{\bf 4}, 153 (1961)];
L. D. Landau and E. M. Lifshitz, {\it Electrodynamics of Continuous Media},
(Pergamon, Oxford, 1960), pp.~368--376.

\bibitem{sabisky} E. S. Sabisky and C. H. Anderson, {\it Phys.\ Rev.} A {\bf 7},
790 (1973).

\bibitem{archive} Julian Schwinger Papers (Collection 371), Department of
Special Collections, University Research Library, University of California,
Los Angeles.

\bibitem{174} J. Schwinger, {\it Lett.\ Math.\ Phys.} {\bf 1}, 43 (1975).

\bibitem{sm} F. Sauer, dissertation, Gottingen, 1962, unpublished;
J. Mehra, dissertation; {\it Acta Physica Austriaca\/} (1964);
{\it Physica} {\bf 37}, 145 (1967).

\bibitem{har} C. M. Hargreaves, {\it Proc.\ Kon.\ Ned. Akad.\
 Wetensch.} B {\bf 68},
231 (1965).

\bibitem{lifcorr} Letter to J. Schwinger from E. M. Lifshitz, dated 27
April 1978.

\bibitem{187} J. Schwinger, L. L. DeRaad, Jr., and K. A. Milton, 
{\it Ann.\ Phys.} (N.Y.) {\bf 115}, 1 (1978).

\bibitem{london} F. London, {\it Z. Phys.} {\bf 63}, 245 (1930).

\bibitem{188} K. A. Milton, L. L. DeRaad, Jr., and J. Schwinger,
{\it Ann.\ Phys.} (N.Y.) {\bf 115}, 388 (1978).

\bibitem{casimir2} H. B. G. Casimir, {\it Physica} {\bf 19}, 846 (1956).

\bibitem{boyer} T. H. Boyer, {\it Phys.\ Rev.} {\bf 174}, 1764 (1968).

\bibitem{davies} B. Davies, {\it J. Math.\ Phys.} {\bf 13}, 1324 (1972).

\bibitem{baldup} R. Balian and B. Duplantier, {\it Ann.\ Phys.} 
(N. Y.) {\bf 112}, 165 (1978).

\bibitem{boyerlet} Letter to L. L. DeRaad, Jr., from T. H. Boyer,
dated 12 May 1978.

\bibitem{balianlet} Letter to J. Schwinger from R. Balian, dated 28
December 1977.

\bibitem{dimcas} C. M. Bender and K. A. Milton, {\it Phys.\ Rev.} D {\bf 50},
6547 (1994); K. A. Milton, {\it Phys.\ Rev.} D {\bf 55}, 4940 (1997).


\bibitem{unruh} P. C. W. Davies, {\it J. Phys.\ A: Gen.\ Phys.} {\bf 8}, 365
(1975);
 W. G. Unruh, {\it Phys.\ Rev.} D {\bf 14}, 870 (1976);
N. D. Birrell and P. C. W. Davies, {\it Quantum Field Theory in
Curved Space\/} (Cambridge University Press, Cambridge, England, 1982).

\bibitem{sigxi} 1986 Monie A. Ferst Symposium and Banquet Honoring
Professor Julian Schwinger, May 20, 1986, Georgia Institute of Technology.

\bibitem{hawking} S. W. Hawking, {\it Nature\/} {\bf 248}, 30 (1974);
{\it Commun.\ Math.\ Phys.} {\bf 43}, 199  (1975).

\bibitem{villa} E-mail message from M. Villasante to K. Milton, September 1998.

\bibitem{kara}  Evangelos Karagiannis,
{\it Radiative Polarization of a Rotating Charge and Construction of
Green's Function}, UCLA Ph.D. Thesis, UMI-90-35228-mc (microfiche), 1990. 66pp. 

\bibitem{arnowitt} Interview with Richard Arnowitt by K. Milton, July 1998.

 \bibitem{fink} Interview with Robert Finkelstein by K. Milton, July 1997.




June 1997.




\bibitem{sono} H. Frenzel and H. Schultes, {\it Z. Phys.\ Chem}.,
 Abt.\ B {\bf 27}, 421 (1934).

\bibitem{putt} B. P. Barber, R. A. Hiller, R. L\"ofstedt, S. J. Putterman,
and K. Weniger, {\it Phys.\ Rep.} {\bf 281}, 65 (1997).


\bibitem{barton} K. Scharnhorst, {\it Phys.\ Lett.} {\bf B236}, 354 (1990);
G. Barton, {\it Phys.\ Lett.} {\bf B237}, 559 (1990); see also
G. Barton and K. Scharnhorst, {\it J. Phys.\ A} {\bf 26}, 2037 (1993).

\bibitem{221} J. Schwinger, {\it Lett.\ Math.\ Phys.} {\bf 24}, 59 (1992).

\bibitem{222} J. Schwinger, {\it Lett.\ Math.\ Phys.} {\bf 24}, 227 (1992).


\bibitem{feynbio} J. Mehra, {\it The Beat of a Different Drum:
The Life and Science of Richard Feynman} (Oxford University Press,
Oxford, 1994), p.~395.

\bibitem{223} J. Schwinger, {\it Proc.\ Natl.\ Acad.\ Sci.\ USA} {\bf 89}, 4091
(1992).
 
 
\bibitem{224} J. Schwinger, {\it Proc.\ Natl.\ Acad.\ Sci.\ USA} {\bf 89}, 11118
(1992).

\bibitem{sagredo} J. Schwinger, {\it Particles, Sources, and Fields}
(Addison-Wesley, Reading, Mass., 1970) p.~241, explicates this Galilean
confusion.
 
\bibitem{kim} K. A. Milton, {\it Ann.\ Phys.} (N.Y.) {\bf 127}, 49 (1980);
 K. A. Milton and Y. J. Ng, {\it Phys.\ Rev.}
 E {\bf 55}, 4207 (1997); {\bf 57}, 5504 (1998).
 
 \bibitem{225} J. Schwinger, {\it Proc.\ Natl.\ Acad.\ Sci.\ USA} {\bf 90}, 958,
2105, 4505, 7285 (1993).

\bibitem{230} J. Schwinger, {\it Proc.\ Natl.\ Acad.\ Sci.\ USA}
{\bf 91}, 6473 (1994).

 \bibitem{brevikandbarton} I. Brevik and V. N. Marachevsky, `Casimir
 Surface Force on a Dilute Dielectric Ball,'
  I. Brevik, V. N. Marachevsky, and K. A. Milton,
 hep-th/9810062; G. Barton, `Perturbative Check on the Casimir
 Energy on a Nondispersive Ball.'


\bibitem{Proc2} K. Milton, contributed paper to this Proceedings.


\bibitem{229} J. Schwinger, `The Greening
of Quantum Field Theory: George and I,' (Nottingham lecture),
published in {\it Julian Schwinger:
The Physicist, the Teacher, and the Man}, ed.~Y. J. Ng (World Scientific,
Singapore, 1996), p.~9.

\bibitem{eberviss}  C. Eberlein, {\it Phys. Rev.\ A} {\bf 53}, 2772 (1996);
{\it Phys.\ Rev.\ Lett.} {\bf 76}, 3842 (1996);
C. E. Carlson, C. Molina-Par\'\i s, J.
P\'erez-Mercader, and
M. Visser, Phys.\ Lett.\ {\bf B395}, 76 (1997); Phys. Rev. D {\bf 56},
1262 (1997);
C. Molina-Par\'\i s and M. Visser,  Phys. Rev. D {\bf 56}, 6629 (1997);  
S. Liberati, M. Visser, F. Belgiorno, and D. W. Sciama.
``Sonoluminescence: Bogolubov Coefficients for the QED Vacuum of a
Collapsing Bubble,'' quant-ph/9805023; S. Liberati, F. Belgiorno, 
M. Visser, and D. W. Sciama, ``Sonoluminescence as a Quantum Vacuum
Effect,'' quant-ph/9805031.

\end{thebibliography}
\end{document}